%% file: HIN-21-013_temp.tex
\pdfoutput=1
\documentclass[11pt,twoside,a4paper,cmspaper,final,collab]{cms-tdr}
\def\svnVersion{94f3de4}\def\svnDate{2024/09/30}\def\cmsCernNoTag{CERN-EP-2023-281}\def\cmsCernDate{\today}\def\cmsMessage{Published in Physical Review Letters as \href{http://dx.doi.org/10.1103/PhysRevLett.133.142301}{\doi{10.1103/PhysRevLett.133.142301}.}}
\begin{document}\cmsNoteHeader{HIN-21-013}

\newcommand{\roots}    {\ensuremath{\sqrt{s}}}
\newcommand{\jT}         {\ensuremath{j_{\mathrm{T}}}\xspace}
\newcommand{\detastar} {\ensuremath{\Delta\eta^{\ast}}\xspace}
\newcommand{\dphistar} {\ensuremath{\Delta\phi^{\ast}}\xspace}
\newcommand{\etastar} {\ensuremath{\eta^{\ast}}\xspace}
\newcommand{\phistar} {\ensuremath{\phi^{\ast}}\xspace}
\newcommand{\vstarTWO} {\ensuremath{v^{\ast}_2}\xspace}
\newcommand{\VstarNdelta} {\ensuremath{V^{\ast}_{\mathrm{n}\Delta}}\xspace}
\newcommand{\VstarONEdelta} {\ensuremath{V^{\ast}_{\text{1}\Delta}}\xspace}
\newcommand{\VstarTWOdelta} {\ensuremath{V^{\ast}_{\text{2}\Delta}}\xspace}
\newcommand{\VstarTHRdelta} {\ensuremath{V^{\ast}_{\text{3}\Delta}}\xspace}
\newcommand{\AonA}  {\ensuremath{\text{AA}}\xspace}
\newcommand{\nchj}{\ensuremath{N_\text{ch}^\mathrm{j}}}
\newcommand{\nchtrg}{\ensuremath{N_{\text{ch}}^{\text{trg}}}}
\newcommand{\bdist}{\ensuremath{B(\detastar,\dphistar)}}
\newcommand{\sdist}{\ensuremath{S(\detastar,\dphistar)}}

\cmsNoteHeader{HIN-21-013}

\title{Observation of enhanced long-range elliptic anisotropies inside high-multiplicity jets in \texorpdfstring{$\Pp\Pp$}{pp} collisions at \texorpdfstring{$\roots=13\TeV$}{sqrt(s) = 13 TeV}}

\date{\today}

\abstract{A search for collective effects inside jets produced in proton-proton collisions is performed via correlation measurements of charged particles using the CMS detector at the CERN LHC. The analysis uses data collected at a center-of-mass energy of $\roots=13\TeV$, corresponding to an integrated luminosity of 138\fbinv. Jets are reconstructed with the anti-$\kt$ algorithm with a distance parameter of 0.8 and are required to have transverse momentum greater than 550\GeV and pseudorapidity $\abs{\eta^{\rm jet}}<1.6$. Two-particle correlations among the charged particles within the jets are studied as functions of the particles' azimuthal angle and pseudorapidity separations ($\dphistar$ and $\detastar$) in a jet coordinate basis, where particles' $\etastar$, $\phistar$ are defined relative to the direction of the jet. The correlation functions are studied in classes of in-jet charged-particle multiplicity up to $\nchj\approx100$. Fourier harmonics are extracted from long-range azimuthal correlation functions to characterize azimuthal anisotropy for $\abs{\detastar}>2$. For low-$\nchj$ jets, the long-range elliptic anisotropic harmonic, $\vstarTWO$, is observed to decrease with $\nchj$. This trend is well described by Monte Carlo event generators. However, a rising trend for $\vstarTWO$ emerges at $\nchj\gtrsim80$, hinting at a possible onset of collective behavior, which is not reproduced by the models tested. This observation yields new insights into the dynamics of jet evolution in the vacuum.}

\hypersetup{
pdfauthor={CMS Collaboration},
pdftitle={Observation of enhanced long-range elliptic anisotropies},
pdfsubject={CMS},
pdfkeywords={CMS, jet, collectivity, 13 TeV, pp}
}

\maketitle

The theory of quantum chromodynamics (QCD) describes the strong interaction among partons (quarks and gluons) carrying color charges. At low energies, the color force is very strong and partons are confined in color-neutral objects (hadrons) --- a property that results from the nonperturbative nature of QCD in this regime.  
In high energy proton collisions,  large momentum transfers between partons inside the colliding protons can result in a collimated spray of hadrons originating from the fragmentation and hadronization of an outgoing parton. This collimated spray is called a ``jet.''~\cite{Sterman:1977wj,CTEQ:1993hwr}.  Energetic jets are produced abundantly at LHC collision energies and can generate large final-state hadron multiplicities (e.g.,  $>100$ charged particles) resulting from a single parton.

Final states containing thousands of hadrons are routinely produced in high-energy nucleus-nucleus ($\AonA$) collisions, 
where a hot medium of deconfined partons known as a 
quark-gluon plasma (QGP) is formed~\cite{Adams:2005ph,Alver:2008gk,Abelev:2009af,Alver:2009id,Muller:2012zq,Busza:2018rrf,CMS:2024sgx}. 
The extreme parton densities realized in these collisions result in strong 
partonic rescatterings which quickly drive the system toward a nearly ideal hydrodynamic limit. As a result, long-range collective 
flow effects have been observed at the BNL RHIC~\cite{Adams:2005ph,Alver:2008gk,Abelev:2009af,Alver:2009id} 
and the CERN LHC~\cite{Chatrchyan:2011eka,Chatrchyan:2012wg,Aamodt:2010pa,ATLAS:2012at,CMS:2013bza}. 
 
It was originally thought that small collision systems such as electron-positron ($\EE$), electron-proton ($\Pe\Pp$), and proton-proton ($\Pp\Pp$) collisions would produce final states that were too small and dilute for secondary partonic rescatterings to drive the system toward thermal equilibrium. Collective hydrodynamic behavior was not expected to play an important role in these final states, notwithstanding some early studies~\cite{david1,david2}. Surprisingly, strong long-range collective correlations, similar to those observed in $\AonA$ collisions, were discovered in the azimuthal distributions of charged particles in the laboratory reference frame of $\Pp\Pp$ collisions having a large final-state multiplicity in the entire event~\cite{Khachatryan:2010gv,Aad:2015gqa,Khachatryan:2015lva,Khachatryan:2016txc}. This raised the question of whether a tiny QGP droplet is created in such conditions~\cite{Dusling:2015gta}. Subsequently, 
similar collective phenomena were observed in 
proton-nucleus ($\Pp$A) collisions ~\cite{CMS:2012qk,alice:2012qe,Aad:2012gla,Aaij:2015qcq,ABELEV:2013wsa,CMS:2013jlh,Khachatryan:2014jra,Khachatryan:2015waa,Aaboud:2017acw,Aaboud:2017blb,PHENIX:2018lia},
and lighter AA systems~\cite{Adamczyk:2015xjc,Adare:2015ctn,Aidala:2017ajz,PHENIX:2018lia}.
Correlation studies in $\EE$~\cite{Badea:2019vey,Belle:2022fvl,alepCorr:2023}, 
$\Pe\Pp$~\cite{ZEUS:2019jya,ZEUS:2021qzg}, photon-proton ($\PGg\Pp$)~\cite{CMS:2022doq} and photonuclear ($\PGg$A)~\cite{ATLAS:2021jhn} systems 
have been limited to relatively low-multiplicity events (less than a total of 30--40 charged particles per event) and have not unambiguously demonstrated such behavior.

\begin{figure}[!ht]
\centering
\includegraphics[width=0.9\columnwidth]{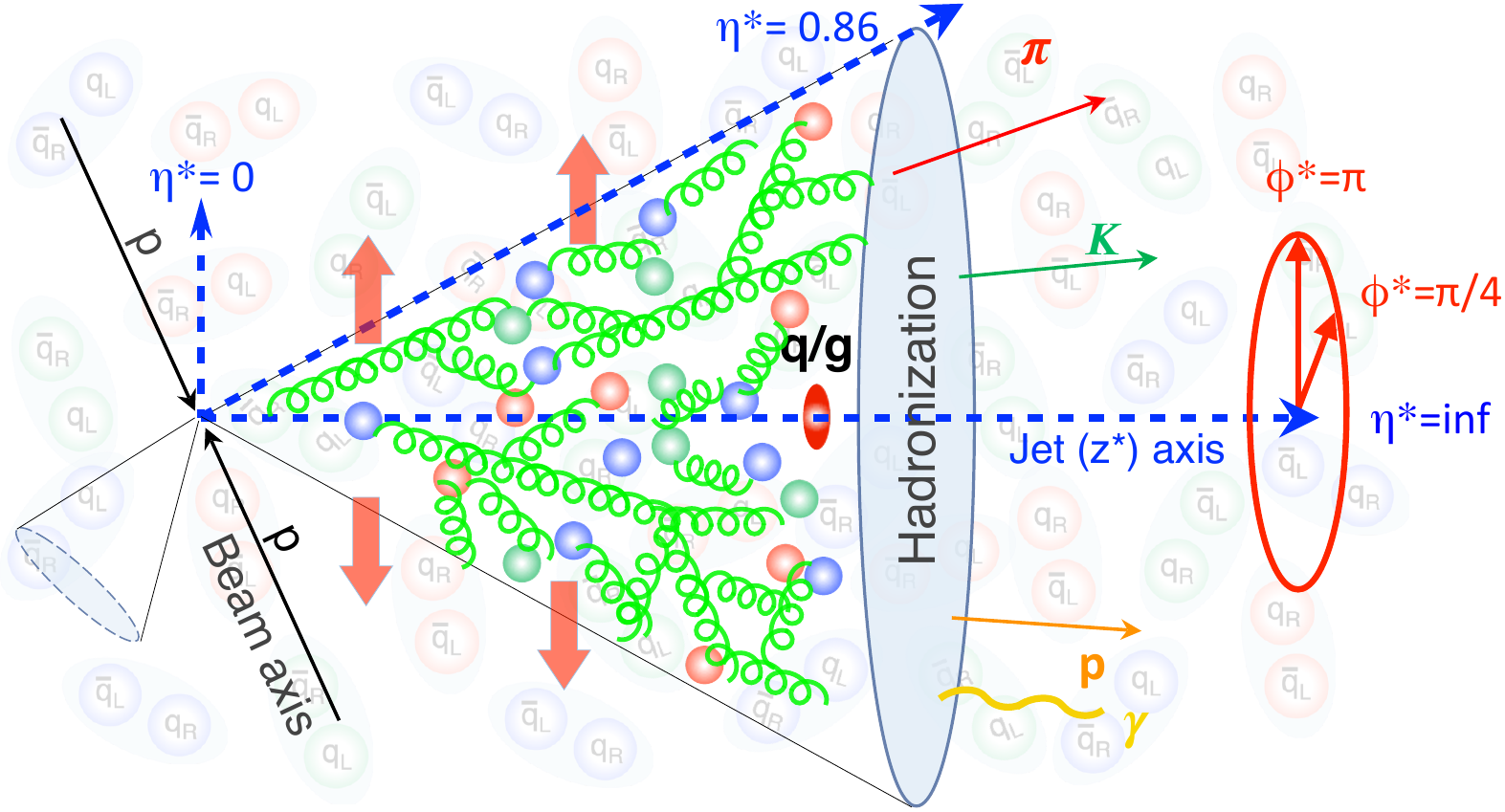}
\caption{An illustration showing the idea of an initially scattered parton evolving to a shower that eventually exhibits collective expansion transverse to the jet axis (represented by red arrow boxes). A jet cone (not to scale) and emerging final state particles are drawn in a coordinate system, denoted as the ``jet basis,'' where the $z$ axis coincides with the jet direction. The redefined pseudorapidity $\etastar$ is shown with key values in dotted blue lines, and the azimuthal angle $\phistar$ is shown with key values in solid red.}
\label{cartoon}
\end{figure}

It is natural to ask what minimum system size is needed for QCD collective effects to develop. Although the dynamics of parton showering inside a jet is theoretically well described by perturbative QCD calculations~\cite{david3,david4}, a possible build-up of collective correlations within the parton constituents of a jet had not been considered. In Ref.~\cite{Baty:2021ugw}, it is postulated that collective effects
can emerge from an initial system as small as an energetic parton that fragments and hadronizes in the vacuum, as illustrated in Fig.~\ref{cartoon}. Motivated by that idea, this Letter presents a 
search for such collective effects inside individual jets (as opposed to full events) produced in $\Pp\Pp$ collisions 
at a center-of-mass energy of $\roots=13\TeV$ using the CMS detector at the LHC. 
Using a coordinate system defined with respect to the jet axis --- a proxy for the direction of the parton initiating the jet --- the two-particle correlation of charged particles of a jet is measured as a function of the in-jet charged particle multiplicity, $\nchj$. Tabulated results are provided in the HEPData record for this analysis~\cite{hepdata}.

The CMS apparatus~\cite{CMS:2008xjf} is a multipurpose, nearly hermetic detector, designed to trigger 
on~\cite{CMS:2020cmk,CMS:2016ngn} and identify electrons, muons, photons, and 
hadrons~\cite{CMS:2020uim,CMS:2018rym,CMS:2014pgm}. The initial triggering is done with the level-1 system, which uses customized hardware to make the rapid online decision whether or not to accept an event and deliver it to the second system, the High Level Trigger (HLT). The HLT uses a large CPU farm to perform optimized online event reconstruction and characterize an event. A global ``particle-flow" algorithm~\cite{CMS:2017yfk} 
reconstructs all individual particles in an event, combining information provided by the all-silicon inner 
tracker and by the crystal electromagnetic and brass-scintillator hadron calorimeters, operating inside a 3.8\unit{T} 
superconducting solenoid, with data from the gas-ionization muon detectors embedded in the flux-return 
yoke outside the solenoid. The reconstructed particles are used to build more complex objects such as jets~\cite{CMS:2018jrd,CMS:2016lmd,CMS:2019ctu,CMS:2016xef}. 

The $\roots=13\TeV$ $\Pp\Pp$ collisions used in this analysis were delivered from 2016-2018 and correspond to an integrated luminosity of 138\fbinv. The data were collected using an online trigger searching for events 
containing anti-\kt jets~\cite{Cacciari:2008gp,Cacciari:2011ma} with distance parameter $\mathrm{R} = 0.8$ having a transverse momentum ($\pt^\text{jet}$) above $500\GeV$.
In the offline analysis, jets were required to have $\pt^\text{jet}>550\GeV$ and pseudorapidity $\abs{\eta^\text{jet}}<1.6$ in the laboratory reference frame.

The generator-level events from two Monte Carlo (MC) models, \PYTHIA 8.306~\cite{Sjostrand:2007gs} with the CP5 tune~\cite{CMS:2019csb} and \SHERPA 2.2~\cite{Sherpa:2019gpd}, are compared with the data. The leading order \PYTHIA 8 uses the Lund String fragmentation model~\cite{Andersson:1983ia}, while the \SHERPA generator uses a cluster fragmentation model~\cite{Winter:2003tt}. Additionally, the \PYTHIA 8 events are input into \GEANTfour~\cite{Agostinelli:2003} to emulate the CMS detector response for calculating correction factors for the tracking efficiency and jet energy scale. The effects of secondary $\Pp\Pp$ interactions within the same or nearby bunch crossings (``pileup'') are incorporated in the simulation.

This analysis is particularly interested in the charged particles of jets. These charged particles are required to have $\abs{\eta}<2.4$ and $\pt>0.3\GeV$ in the laboratory reference frame. Additional, they must have a \pt uncertainty of $<10$\%, and a distance of closest approach significance with respect to the primary vertex of at most three standard deviations ($\sigma$)~\cite{CMS:2014pgm}. The PUPPI algorithm~\cite{Sirunyan:2020foa,Bertolini:2014bba} is used to mitigate the effect of pileup at the reconstructed particle level, using local shape information, event pileup properties, and tracking information. Jets are classified into different classes based on the number of charged particles of the jet passing these selections and PUPPI subtraction before correcting for detector effects. 

Jet energy corrections are derived from MC simulation studies so that the average measured energy of jets equals that of generator-level jets.
Additional selection criteria are applied to each jet to remove jets potentially dominated by instrumental effects or reconstruction failures. The jet energy resolution is typically 10\% at 100\GeV, 
and 5\% at 1\TeV~\cite{CMS:2016lmd}.

The two-particle correlation analysis is similar to that
employed in Ref.~\cite{CMS:2012qk}, except that the momentum vectors of all charged particles 
of a jet are defined in the ``jet basis'', as illustrated in Fig.~\ref{cartoon}. A unique jet reference coordinate basis is defined for every jet such that the $z$-axis is aligned with the direction of the jet momentum~\cite{Baty:2021ugw}. Momentum vectors of charged particles are
redefined in this new basis, $\vec{p^\ast} = (\jT,\etastar,\phistar)$. Here, $\jT$ is the particle \pt
with respect to the jet axis. The symbols $\etastar$ and $\phistar$ are the pseudorapidity and azimuthal angle coordinates with respect to the jet axis. Therefore, $\etastar= 0$ and $\infty$ correspond to vectors that are perpendicular and parallel to the jet axis, as illustrated in Fig.~\ref{cartoon}. In this system, $\etastar= 0.86$ is the approximate boundary of the anti-$\kt$
jet. The origin of the $\phi^*$ coordinate is defined as $\hat{p}_{\text{jet}}\times(\hat{p}_{\text{jet}}\times\hat{z})$, where $\hat{p}_{\text{jet}}$ is a unit vector along the jet momentum and $\hat{z}$ is the direction of the $+z$ axis. The choice of $\phi^*=0$ is irrelevant to the relative $\phi^*$ between two particles. 
The azimuthal coordinates of tracks that are approximately parallel to the direction of the jet axis ($\etastar > 5$) are sensitive to small changes in the jet axis direction caused by resolution effects. These tracks are excluded in the subsequent analysis. Reconstructed particles in the event that are not clustered into the jet of interest are not considered in the analysis. For each jet with $\pt^{\rm jet}>550\GeV$, the two-dimensional (2D) angular correlation function is calculated using charged particles from the jet as follows:
\begin{linenomath}
\begin{equation}
\frac{1}{\nchtrg}\dd{^2N^{\text{pair}}}{\detastar \rd\dphistar}=B(0,0)\frac{\sdist}{\bdist},
\label{maineq1}
\end{equation}
\end{linenomath}
\noindent where $\detastar$ and $\dphistar$ are the relative separation in pseudorapidity and azimuthal angle in the jet basis between a pair of charged particles selected from a given $\jT$ range. The functions $\sdist$ and $\bdist$ represent the signal and combinatorial distributions, respectively: 
\begin{linenomath}
\begin{equation}
\sdist= \frac{1}{\nchtrg}\dd{^2N_{\text{ch}}^{\text{sig}}}{\detastar\rd\dphistar},
\end{equation}
\noindent and
\begin{equation}
\bdist= \frac{1}{\nchtrg}\dd{^2N_{\text{ch}}^{\text{combin}}}{\detastar\rd\dphistar},
\end{equation}
\end{linenomath}
\noindent where $N_{\text{ch}}^{\text{sig}}$ and $N_{\text{ch}}^{\text{combin}}$ are the numbers of signal and combinatorial pairs, respectively, and $N^{\text{trg}}_{\text{ch}}$ is the number of particles within a specific $\jT$ range used to calculate the correlation functions. 
The measurement is performed for each class of $\nchj$, which is corrected for detector effects (a tracking efficiency of about 90\%). 
The $\sdist$ distribution is calculated with pairs of charged particles from each unique jet individually, and then averaged over all jets. The combinatorial distribution serves as both a reference and a correction to the pair acceptance due to the limited $\etastar$ range. To construct the $\bdist$ distribution, a two-dimensional (2D) single-particle $\etastar$--$\phistar$ distribution for charged particles of all jets within the $\nchj$ range chosen for the signal distribution is first obtained.
Pairs of $(\etastar,\phistar)$  points are then randomly selected from this distribution to construct $\bdist$ such that no intrinsic correlations other than the pair acceptance effect are present in $\bdist$. Correlations present in the signal distribution that are related to single-particle distributions
or detector effects are canceled by the $B(0,0)/\bdist$ term in Eq. (\ref{maineq1}). 
The quantities $\detastar$ and $\dphistar$ are always taken to be positive and used to
fill one quadrant of the $(\detastar,\dphistar)$ histograms with the other three quadrants filled by reflection.

The resulting 2D distribution can be further studied by 
integrating over $\abs{\detastar}>2$ and decomposing into a one-dimensional (1D) Fourier series, as done in previous analyses~\cite{Khachatryan:2010gv,Khachatryan:2015lva,Khachatryan:2016txc,CMS:2012qk,CMS:2013jlh,Khachatryan:2014jra,Khachatryan:2015waa}:
\begin{linenomath}
\begin{equation}
\frac{1}{\nchtrg} \dd{N^{\text{pair}}}{\dphistar} \propto 1+2\sum_{n=1}^{\infty}\VstarNdelta\cos(\text{n}\dphistar). 
\label{eq:1}
\end{equation}
\end{linenomath}
By applying $\abs{\detastar} > 2$, short-range few-body correlations, such as resonance and cluster decays~\cite{Khachatryan:2010gv}, are excluded. 
Of particular interest is the second Fourier component, which is associated with elliptic anisotropies. The elliptic anisotropy coefficient, $\vstarTWO$, is the main observable of interest and is related to the two-particle Fourier coefficient by assumed factorization as $\vstarTWO=\sqrt{\VstarTWOdelta}$~\cite{CMS:2013jlh,Khachatryan:2016txc}.

The primary sources of systematic uncertainty come from the jet axis pointing resolution, the $\pt^\text{jet}$ resolution, residual pileup effects, and the tracking efficiency. These four sources are added together in quadrature to calculate the total systematic uncertainty for each $\nchj$ class.
To minimize statistical fluctuations, the four highest jet multiplicity classes examined are combined together when evaluating systematic effects.
The effect of the jet pointing resolution is studied by smearing each jet axis in the 2D $\eta$-$\phi$ plane according to the resolution calculated in MC samples. 
The coordinate transformation of particles' $\jT$, $\etastar$, and $\phistar$ into the jet basis is recalculated with the new smeared jet axis, and the analysis procedure is repeated. 
The difference of $\VstarNdelta$ values before and after smearing is taken as the systematic uncertainty. The impact of this effect on $\VstarNdelta$ is between 0.01 and 0.04 (depending on the $\jT$ range examined) at lower $\nchj$, where particles inside a jet tend to have a narrower angular distribution and therefore larger changes to their jet basis kinematic values for a given jet axis variation. On the other hand, this uncertainty is $<$0.001 at higher $\nchj$ because the particles of these jets tend to have a wider angular separation from the jet axis on average.
A similar smearing procedure found the uncertainty in $\VstarNdelta$ coming from the $\pt^\text{jet}$ resolution to range from negligible to 0.002.
The systematic uncertainty from residual pileup effects is estimated by first splitting the data into two roughly equal subsets based on the number of reconstructed vertices in each event. The resulting high and low vertex number subsets were then treated independently and the full analysis procedure was carried out on both. The root mean square of the deviations in measured observables of these two subsets from the nominal result is taken as the uncertainty. For $\VstarTWOdelta$, this uncertainty ranges from 0.001--0.003. Additional pileup studies varying PUPPI selection requirements were found to have a negligible impact. 
Potential contamination of collective effects from the underlying event was investigated by injecting a sizable lab frame $v_2$ signal, and is found to have no impact on the jet frame $\vstarTWO$.
The systematic uncertainty resulting from the tracking efficiency correction is evaluated by applying tighter and looser restrictions on the relative $\pt$ uncertainty and track vertex association criteria, repeating the analysis procedure, and calculating the root mean square of the deviations of these variations from the nominal case. The uncertainty in $\VstarTWOdelta$ from this source is less than 0.003.

\begin{figure}[th]
\centering
\includegraphics[width=0.98\columnwidth]{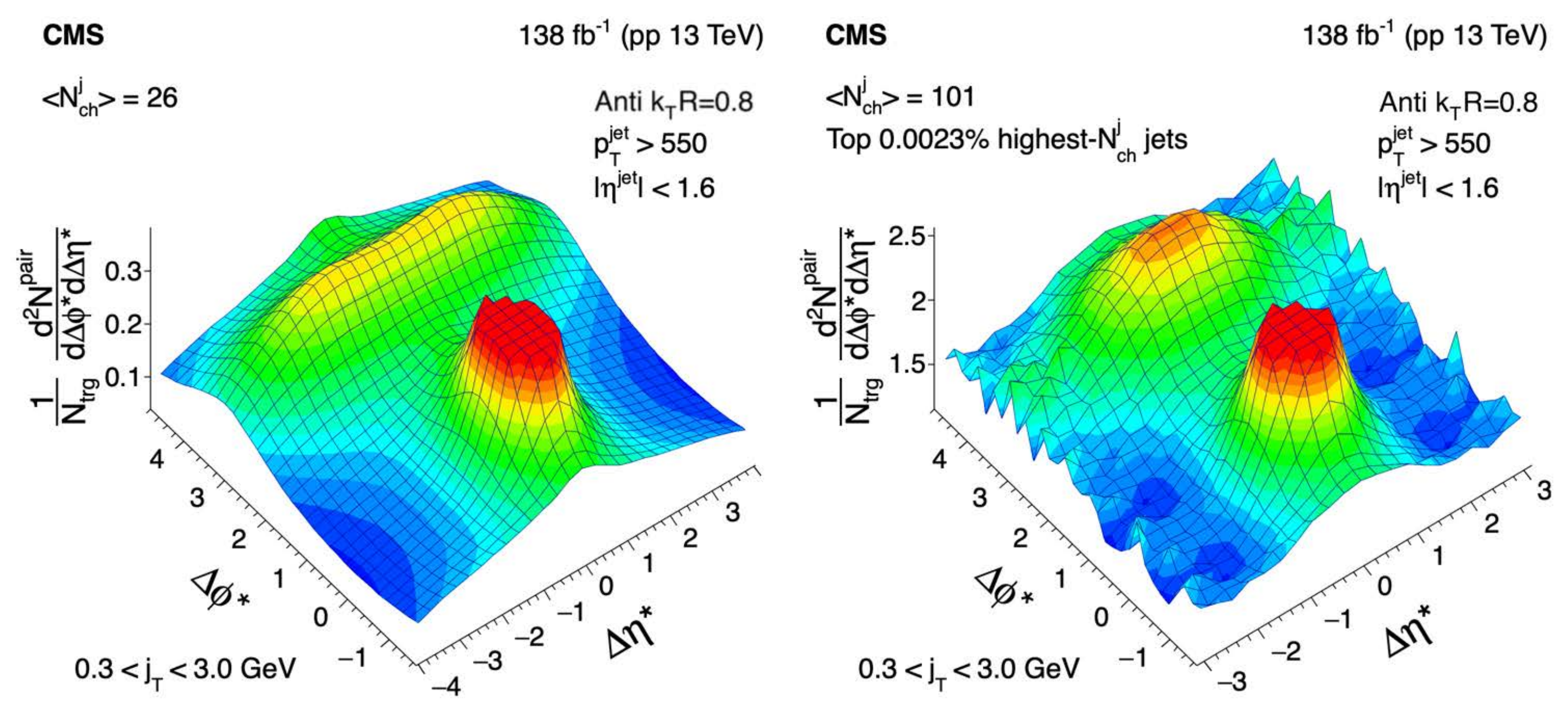}
 \caption{Results for 2D two-particle angular correlation functions for particle $0.3< \jT <3.0\GeV$ from inclusive (left) and the highest (right) $\nchj$ jets, for anti-$\kt$ $\mathrm{R}=0.8$ algorithm with $\pt^{\rm jet} > 550\GeV$ and $\abs{\eta^{\rm jet}} < 1.6$.}
\label{main2d}
 \end{figure}

Figure~\ref{main2d} shows an example of 2D two-particle angular correlation functions in the jet basis for inclusive and high-$\nchj$ jets 
for charged particles in the range $0.3<\jT<3.0\GeV$. The average $\nchj$ value for the two classes of jets shown, $\langle\nchj\rangle$, is 26 and 101.
The high-$\nchj$ jet class corresponds to a fraction of only $2\ten{-5}$ of all jets with $\pt^\text{jet}>550\GeV$.
The central peak at $(\detastar,\dphistar) = (0,0)$, truncated for better visualization, is the result of short-range correlations 
from the parton shower and hadronization. The far-side ridge at $\dphistar \approx \pi$ is mostly related to back-to-back particle 
production and conservation of momentum. These prominent features have also been found in laboratory-frame analyses of $\Pp\Pp$ collisions~\cite{Khachatryan:2010gv}, where they can also be reproduced using MC simulations for both low- and high-$\nchj$ jets. This indicates that the dynamics of bulk hadron production in the jet fragmentation 
process may share similarities to those of a hadron-hadron 
collision process.
Moreover, a feature commonly observed in AA collisions is the near-side enhancement at $\dphistar \approx 0$ over a long range in $\detastar$, commonly known as the near-side ``ridge''. The $\nchj$ reached in the single jet system of this work is comparable to the event multiplicity of $\Pp\Pp$ collisions where a near-side ridge was first observed using a laboratory-frame analysis~\cite{Khachatryan:2010gv}.
There appears to be some indication of a near-side ridge in the high-$\nchj$ plot in Fig.~\ref{main2d}, however, this feature is less prominent than the ridges observed in $\Pp\Pp$ and pA collisions. There is no corresponding near-side enhancement visible in the high-$\nchj$ 2D distributions from either \PYTHIA 8 or \SHERPA.
 
\begin{figure}[!h]
\centering
\includegraphics[width=0.47\textwidth]{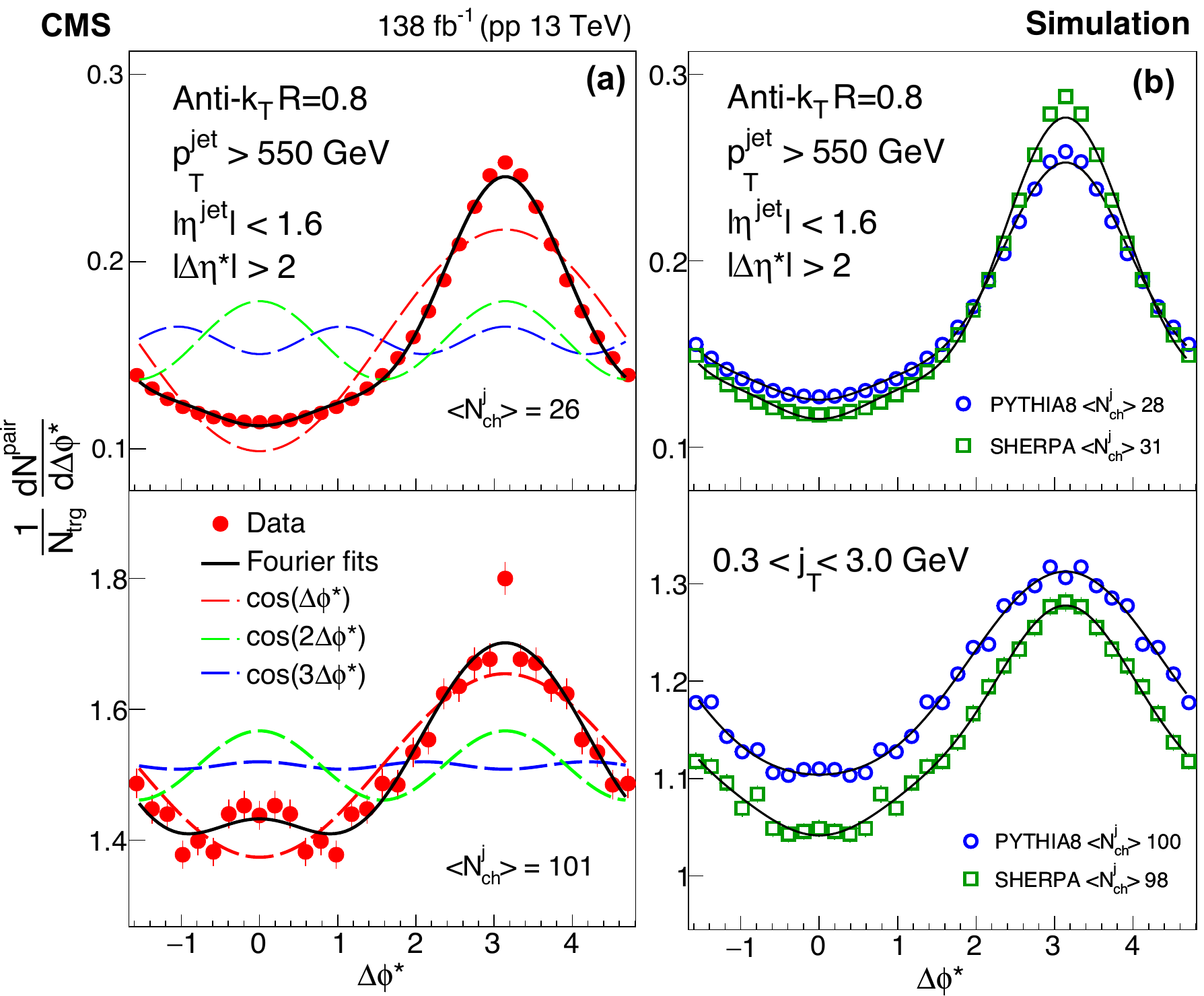}
\includegraphics[width=0.47\textwidth]{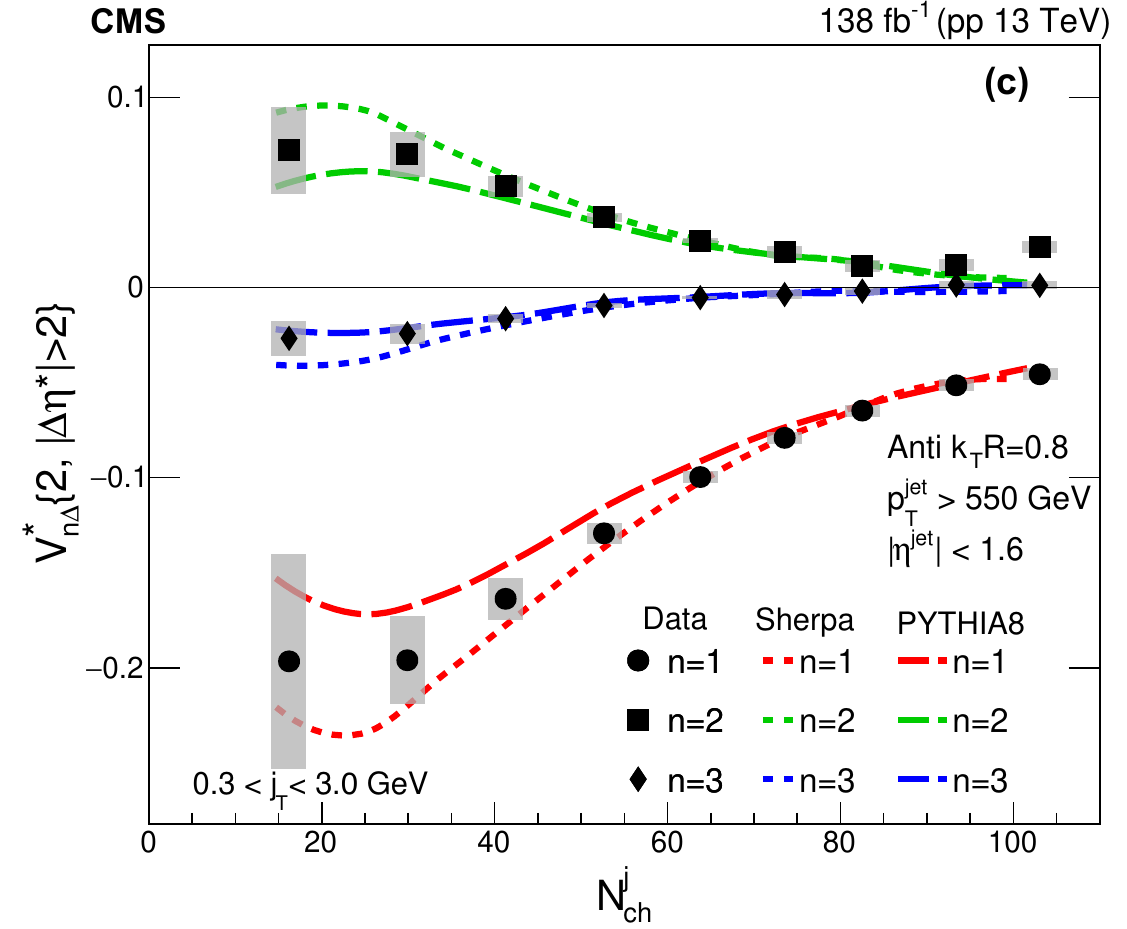}
\caption{Examples of two-particle angular correlations projected onto 1D $\dphistar$ for $\abs{\detastar}>2$. The panels show (a) data and (b) \PYTHIA 8 and \SHERPA for inclusive $\nchj$ (upper) and high-$\nchj$ (lower) jet selections. Data and both MC models are compared in (c) with a continuous evolution of extracted two-particle Fourier coefficients $\VstarNdelta$ as a function of $\nchj$. Vertical bars on data points indicate statistical uncertainty while shaded boxes represent systematic uncertainties. Projections are symmetrized about $\dphistar=0$ and $\pi$.}
\label{fig:main1d}
\end{figure}

The 1D $\dphistar$ correlation functions extracted using Eq.~\eqref{eq:1} and averaged over $\abs{\detastar} > 2$ are shown for data and MC in Fig.~\ref{fig:main1d}(a) and (b), respectively, for particles with $0.3<\jT<3.0\GeV$ from jets in two $\nchj$ classes. 
For high- and inclusive $\nchj$ classes in data, \PYTHIA 8 and \SHERPA, strong away-side correlations observed are consistent with dominant contributions of back-to-back momentum conservation effects. For inclusive $\nchj$ jets, the near-side at $\dphistar \approx 0 $ clearly shows a minimum. However, for the class of highest-$\nchj$ jets studied in data, with $\langle\nchj\rangle\approx100$, (Fig.~\ref{fig:main1d}(a) lower) an indication of a near-side enhancement is seen, which is less obvious or possibly absent from \PYTHIA 8 or \SHERPA events having comparable $\nchj$ (Fig.~\ref{fig:main1d}(b) lower).  
The Fourier fits used to extract $\VstarNdelta$ are also shown and are dominated by a negative $\VstarONEdelta$ component. The addition of more Fourier terms has little impact on the first three Fourier coefficients. The significant deviation from the fit at $\dphistar \approx 0 $ in Fig.~\ref{fig:main1d}(a) lower is attributed to a local statistical fluctuation, as similar deviations are not observed in any other $\nchj$ class.

\begin{figure}[!ht]
\centering
\includegraphics[width=0.4\textwidth]{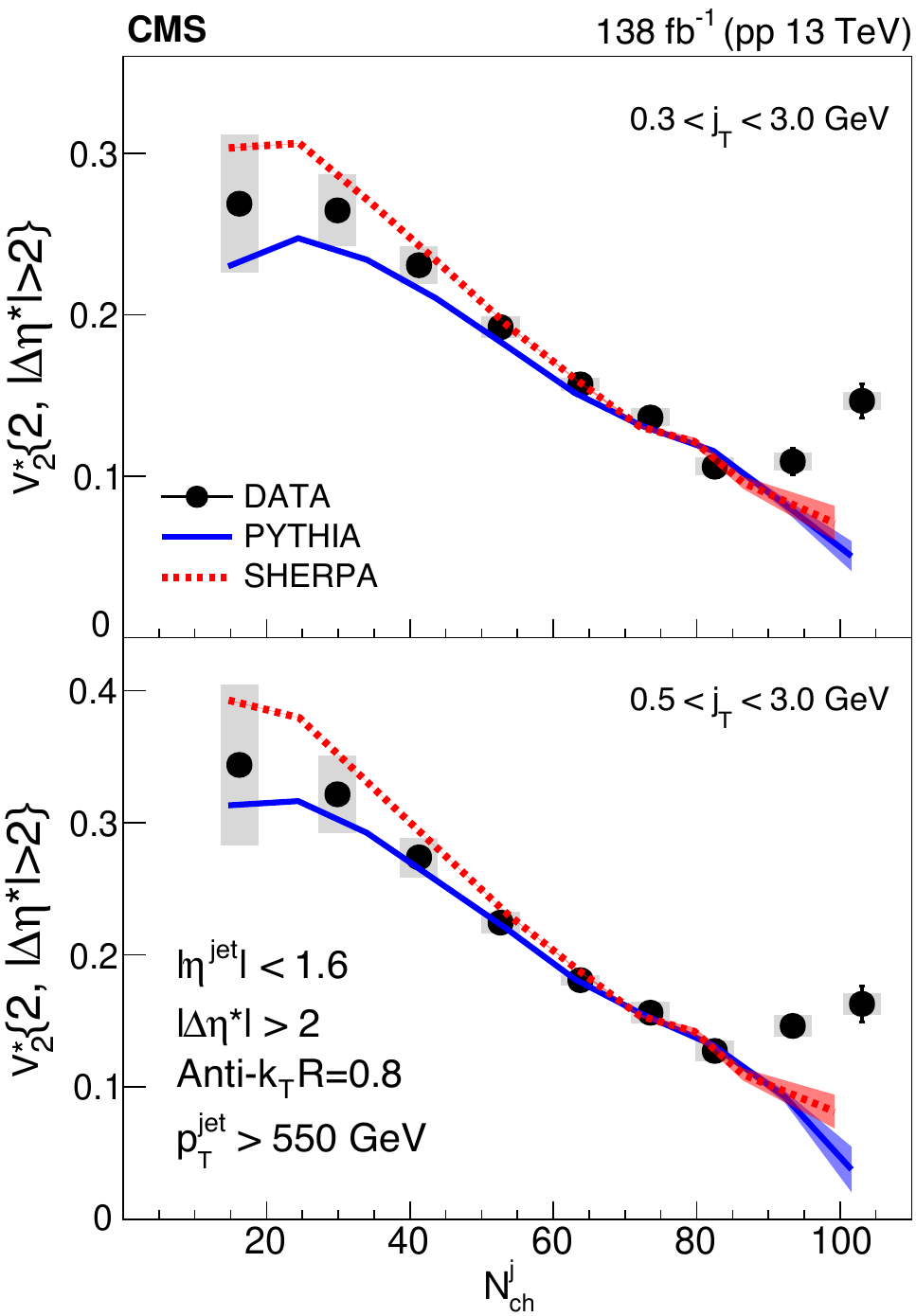}
\caption{The elliptic anisotropies $\vstarTWO$, obtained from two-particle correlations as a function of $\nchj$, for anti-$\kt$ $\mathrm{R}=0.8$ jets with $\pt^{\rm jet}>550\GeV$ and $\abs{\eta^{\rm jet}}<1.6$ in $\Pp\Pp$ collisions at $13\TeV$ from data, \PYTHIA~8, and \SHERPA. Vertical bars on data points indicate statistical uncertainty, while shaded boxes represent systematic uncertainties. The shaded envelope around the MC curves show statistical uncertainty.}
\label{vnevol}
\end{figure}
The extracted two-particle Fourier coefficients for the first three harmonics $\VstarNdelta$, as a function of $\nchj$, are shown in Fig.~\ref{fig:main1d}(c). Data points are placed at the average $\nchj$ value of each jet class for the horizontal axis. Over the full $\nchj$ range, the odd-order harmonics, $\VstarONEdelta$ and $\VstarTHRdelta$, are negative, while $\VstarTWOdelta$ is positive. The magnitudes of all harmonics tend to decrease as $\nchj$ increases. The contribution of few-body correlations to the two-particle Fourier coefficient is expected to diminish as $\nchj$ increases~\cite{Borghini:2001vi}. These features are consistent with back-to-back correlations, as observed in laboratory-frame analyses, that are not related to collective effects. Both MC generators are generally successful in describing the experimental data for all three Fourier harmonics over a wide $\nchj$ range. There appears to be a slight deviation in $\VstarTWOdelta$ between data and simulation.
Figure~\ref{vnevol} shows the elliptic anisotropies $\vstarTWO=\sqrt{\VstarTWOdelta}$, in the jet basis, as a function of $\nchj$ inside the jet. To investigate possible $\jT$ dependence of observed signals, particles from 0.3--3.0\GeV (Fig.~\ref{vnevol} upper) as well as 0.5--3.0\GeV (Fig.~\ref{vnevol} lower) are examined. Again, the MC simulation is generally successful at describing the data over a wide $\nchj$ range in both $\jT$ ranges. For jets at $\nchj>80$, however, the value $\vstarTWO$ no longer diminishes monotonically with increasing $\nchj$. Instead, the data start to show a steady increase with $\nchj$. 
The non-monotonic dependence of $\vstarTWO$ versus $\nchj$ is not expected if few-body processes are the dominant sources of the observed correlations, as in either \PYTHIA 8 or \SHERPA, and may indicate an onset of novel QCD phenomena related to nonperturbative dynamics of a parton fragmenting in the vacuum. These phenomena could include the emergence of collective effects possibly driven by final-state rescatterings, as suggested in Ref.~\cite{Baty:2021ugw}. 
Further experimental and theoretical inputs, including more $\jT$-differential studies with larger data samples, are needed to investigate the physical origin of the observed enhancement.

In summary, the first search for long-range near-side correlations and quantum chromodynamics (QCD) collective effects in jets produced in $\roots=13\TeV$ proton-proton collisions is presented. The measurement is performed using charged particles from individual jets, after their kinematic variables have been calculated in a coordinate basis having the $z$-axis coinciding with the jet direction. Two-particle correlations are studied as a function of the number of charged particles in the jet, $\nchj$. The first three Fourier harmonics of long-range azimuthal correlations are extracted
and compared with those calculated using the \PYTHIA 8 and \SHERPA Monte Carlo (MC) event generators which model the jet fragmentation process. While the data and MC predictions are in good agreement for particle correlations inside jets with $\nchj<80$, the extracted long-range elliptic azimuthal anisotropy $\vstarTWO$ shows a distinct increase in data for $\nchj\gtrsim80$, hinting at a possible onset of collective behavior, which is not reproduced by the MC simulations. 

\begin{acknowledgments}
We congratulate our colleagues in the CERN accelerator departments for the excellent performance of the LHC and thank the technical and administrative staffs at CERN and at other CMS institutes for their contributions to the success of the CMS effort. In addition, we gratefully acknowledge the computing centers and personnel of the Worldwide LHC Computing Grid and other centers for delivering so effectively the computing infrastructure essential to our analyses. Finally, we acknowledge the enduring support for the construction and operation of the LHC, the CMS detector, and the supporting computing infrastructure provided by the following funding agencies: SC (Armenia), BMBWF and FWF (Austria); FNRS and FWO (Belgium); CNPq, CAPES, FAPERJ, FAPERGS, and FAPESP (Brazil); MES and BNSF (Bulgaria); CERN; CAS, MoST, and NSFC (China); MINCIENCIAS (Colombia); MSES and CSF (Croatia); RIF (Cyprus); SENESCYT (Ecuador); MoER, ERC PUT and ERDF (Estonia); Academy of Finland, MEC, and HIP (Finland); CEA and CNRS/IN2P3 (France); SRNSF (Georgia); BMBF, DFG, and HGF (Germany); GSRI (Greece); NKFIH (Hungary); DAE and DST (India); IPM (Iran); SFI (Ireland); INFN (Italy); MSIP and NRF (Republic of Korea); MES (Latvia); LAS (Lithuania); MOE and UM (Malaysia); BUAP, CINVESTAV, CONACYT, LNS, SEP, and UASLP-FAI (Mexico); MOS (Montenegro); MBIE (New Zealand); PAEC (Pakistan); MES and NSC (Poland); FCT (Portugal); MESTD (Serbia); MCIN/AEI and PCTI (Spain); MOSTR (Sri Lanka); Swiss Funding Agencies (Switzerland); MST (Taipei); MHESI and NSTDA (Thailand); TUBITAK and TENMAK (Turkey); NASU (Ukraine); STFC (United Kingdom); DOE and NSF (USA).
\end{acknowledgments}

\bibliography{auto_generated}

\cleardoublepage \appendix\section{The CMS Collaboration \label{app:collab}}\begin{sloppypar}\hyphenpenalty=5000\widowpenalty=500\clubpenalty=5000\input{HIN-21-013-public-authorlist.tex}\end{sloppypar}
\end{document}

%% file: HIN-21-013-public-authorlist.tex
\cmsinstitute{Yerevan Physics Institute, Yerevan, Armenia}
{\tolerance=6000
A.~Hayrapetyan, A.~Tumasyan\cmsAuthorMark{1}\cmsorcid{0009-0000-0684-6742}
\par}
\cmsinstitute{Institut f\"{u}r Hochenergiephysik, Vienna, Austria}
{\tolerance=6000
W.~Adam\cmsorcid{0000-0001-9099-4341}, J.W.~Andrejkovic, T.~Bergauer\cmsorcid{0000-0002-5786-0293}, S.~Chatterjee\cmsorcid{0000-0003-2660-0349}, K.~Damanakis\cmsorcid{0000-0001-5389-2872}, M.~Dragicevic\cmsorcid{0000-0003-1967-6783}, P.S.~Hussain\cmsorcid{0000-0002-4825-5278}, M.~Jeitler\cmsAuthorMark{2}\cmsorcid{0000-0002-5141-9560}, N.~Krammer\cmsorcid{0000-0002-0548-0985}, A.~Li\cmsorcid{0000-0002-4547-116X}, D.~Liko\cmsorcid{0000-0002-3380-473X}, I.~Mikulec\cmsorcid{0000-0003-0385-2746}, J.~Schieck\cmsAuthorMark{2}\cmsorcid{0000-0002-1058-8093}, R.~Sch\"{o}fbeck\cmsorcid{0000-0002-2332-8784}, D.~Schwarz\cmsorcid{0000-0002-3821-7331}, M.~Sonawane\cmsorcid{0000-0003-0510-7010}, S.~Templ\cmsorcid{0000-0003-3137-5692}, W.~Waltenberger\cmsorcid{0000-0002-6215-7228}, C.-E.~Wulz\cmsAuthorMark{2}\cmsorcid{0000-0001-9226-5812}
% ...
\par}
\vskip\cmsinstskip
\dag:~Deceased\\
$^{1}$Also at Yerevan State University, Yerevan, Armenia\\
$^{2}$Also at TU Wien, Vienna, Austria\\
% ...
